\documentclass[%
 reprint,
nofootinbib,
 amsmath,amssymb,
 aps,
]{revtex4-2}
\setcitestyle{authoryear,round} 

\usepackage{graphicx}
\usepackage{dcolumn}
\usepackage{bm}
\usepackage{verbatim}
\usepackage{makecell}

\begin{document}

\preprint{APS/123-QED}

\title{Improvised Nuclear Weapons with 60\%-Enriched Uranium}

\author{Matt Caplan}
 \email{mecapl1@ilstu.edu}
\affiliation{%
 Illinois State University, Department of Physics, Normal IL 61761
}%

\date{\today}

\begin{abstract}
In this work we show that as little as 40 kg of 60\%-enriched uranium can be used to build a crude nuclear weapon with a kiloton yield. While too large to fit on a missile, such a weapon could be delivered by shipping container. 
This analysis is motivated by the June 2025 Israeli and US attacks on Iran, especially the bombings of the nuclear facilities at Natanz, Fordow, and Isfahan.
The Iranian stockpile of approximately 408 kg of 60\%-enriched uranium is, at the time of writing, inaccessible to IAEA inspectors and stored in secret. The rapid clandestine relocation of this material in June 2025 creates an opportunity for aspiring nuclear terrorists to divert an amount that could be used in the construction of an improvised gun-type nuclear weapon in the style of Little Boy.  
\end{abstract}

\maketitle


\section{\label{sec:intro}Introduction}

Following the June 13, 2025 Israeli surprise attack on Iran, that included air strikes on the Natanz Nuclear Facility and assassinations of thirty generals and nine nuclear scientists, increased activity was observed by satellite at the Fordow Enrichment Facility in Iran. 
In anticipation of further attacks on Fordow and Isfahan, an attack that was telegraphed by US rhetoric and that would eventually come on June 22, Iran likely relocated some or most of its stockpile of 60\%-enriched uranium to other sites. Iranian regime officials have claimed that materials including centrifuges and enriched uranium were removed to secret locations \citep[\textit{e.g.}][reporting in Reuters]{murphy2025us}. Though the fog of war shrouds the reality at this time, we take these claims at face value in this work and consider implications for global security.

Without IAEA access to verify the state of the Iranian stockpile, there is now considerable risk that fissile material could be diverted by an aspiring nuclear terrorist and used to build an improvised nuclear device. Before losing access to Iranian facilities following the aforementioned attacks, the IAEA reported that the Iranian stockpile of 60\%-enriched uranium had reached 408 kg, a number now widely reported in the media. Because 90\%-enrichment is canonically referred to as `weapons grade,' the IAEA numbers have lead to a great deal of confusion both among the public and the arms control community. Popular media has reported widely that this uranium is not usable in a weapon, while the arms control community has attempted to correct this misunderstanding \citep[see, \textit{e.g.}][]{lyman2025iran}. 

 \begin{figure*}[t!]
\includegraphics[trim={0 0 0 0},clip,width=0.99
\textwidth]{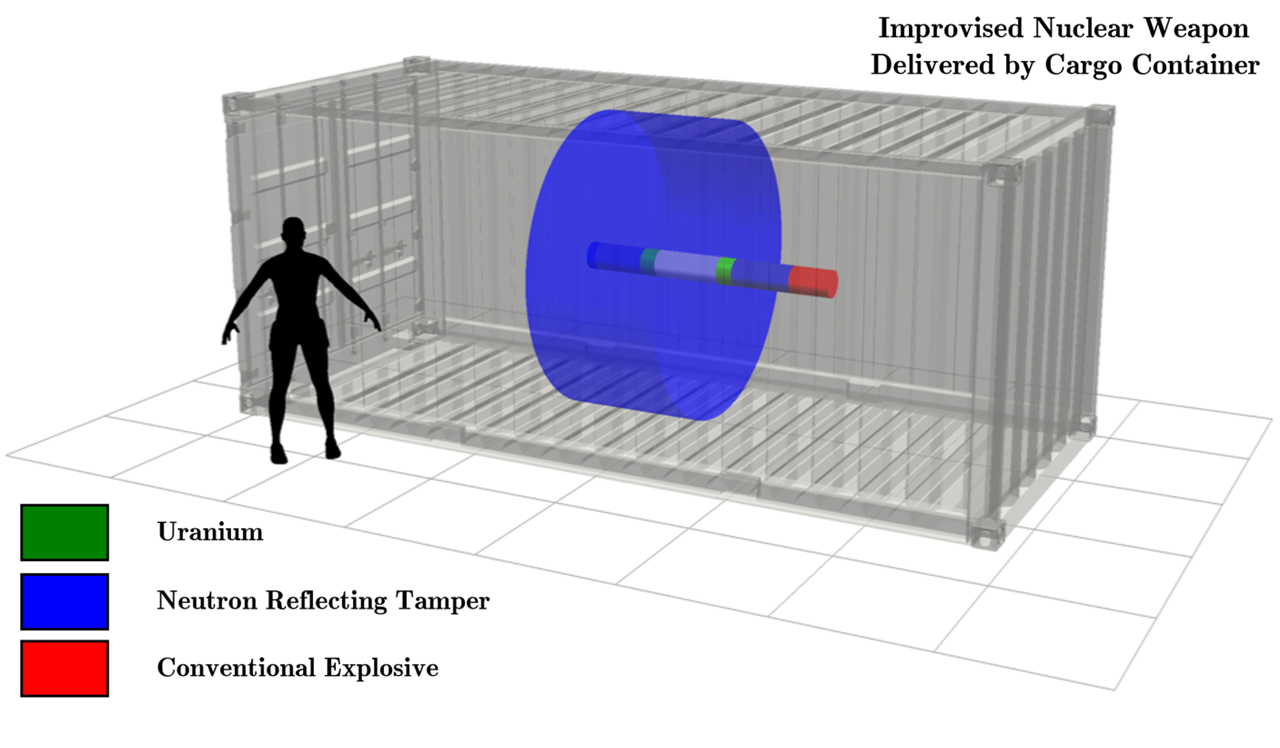}
\caption{\label{fig:schematic} Schematic diagram of a gun-type nuclear weapon using 60\%-enriched uranium (green) deliverable by cargo container, with a mass of 10 to 20 tonnes. The bulk of the weight is given over to the neutron reflecting tamper of tungsten carbide and steel (blue).  Detonation is initiated when the two subcritical uranium masses are rapidly combined into a supercritical assembly inside the tamper by detonating a conventional explosive (red) at the end of a tube. This is effectively a rescaled version of the Little Boy bomb.}
\end{figure*}

There has been much discussion about whether Iran could build a nuclear weapon with this material, but just as important is whether a less well-resourced sub-state or non-state actor could weaponize a small amount of stolen uranium. We argue that an aspiring nuclear terrorist can build a improvised nuclear weapon following the simple design of the gun-type Little Boy bomb dropped on Hiroshima using stolen uranium without further enrichment.\footnote{Papers by \cite{reed2014terrorist}, \cite{friedman2014scenario}, and \cite{mark1986terrorists} consider similar scenarios and are helpful as companions to this work.} A simple schematic of the weapon that will be considered is given in Fig. \ref{fig:schematic}. We will show a high yield is achievable with enormous tampers around a gun-type assembly. A gun-type bomb was famously not tested by the Manhattan Project before being detonated over Hiroshima, as their fissile material was limited and the designers were so confident it would work that there was no need to test, an apt analog for a nuclear terrorist. 

Our work develops the \textsc{FissionBomb} code of \cite{reed2010fission} for lower enrichments by estimating effective isotope-averaged cross sections.
Critical mass considerations for 60\%-enriched uranium are presented in Sec. \ref{sec:crit}. Simulations of the fission chain reactions and weapon yields are given in Sec. \ref{sec:bomb} and we conclude in Sec. \ref{sec:sum}.

\section{Critical Masses at 60\%-Enrichment}
\label{sec:crit}

In this section, we present brief estimates of the bare critical mass and tamped critical mass of uranium, and argue that 60\%-enriched uranium is usable for an improvised nuclear weapon in relatively modest amounts. In Tab. \ref{tab:1} we give our nuclear physics parameters.

\begin{table}[h]
\begin{center}
\vspace{1mm}  
\resizebox{0.5\linewidth}{!}{
\begin{tabular}{ccc}   
\hline \hline   $\phantom{2}^{\phantom{2}^{\phantom{2}}}$
        & $^{235}\textrm{U}$ &  $^{238}\textrm{U}$ \\ 
\hline \hline
$A$                          & 235.05                            & 238.05            \\
$\sigma_f$                          & 1.24                            & 0.100            \\
$\sigma_{el}$                       & 3.557                           & 4.827            \\
$\sigma_{in}$                       & 1.926                           & 2.541            \\
$\sigma_{tr}$                       & 4.52                            & 6.098             \\
$\sigma_a$                          & 0.082                           & 0.07             \\
$\nu$                               & 2.57                            & 2.28           \\ \hline\hline
        & \multicolumn{2}{c}{ 60\%-Enrichment}                   \\ \hline 
$A_{\rm eff}$                          & \multicolumn{2}{c}{ 236.25}             \\
$\sigma_{f,\rm eff}$                          & \multicolumn{2}{c}{ 0.784}             \\
$\sigma_{tr,\rm eff}$                       & \multicolumn{2}{c}{ 5.151}              \\
$\nu_{\rm eff} $   &  \multicolumn{2}{c}{ 2.555}                \\
$\rho $   &  \multicolumn{2}{c}{ $19.05 \, \rm g \, cm^{-3}$}                \\
\hline \hline
\end{tabular}}
\vspace{1mm}  
\end{center}
\caption{\label{tab:1} Nuclear physics constants. Cross sections in barns (1 b = $10^{-24} \, \rm cm^2$). See text for a discussion of $\sigma_f^{238\rm{U}}$. Table adapted from Lamoreaux.}
\end{table}

To compare fissile materials in terms of their suitability for nuclear weapons, analysts often refer to the bare critical mass. 
This is defined as the minimum amount of fissile material shaped into an unreflected sphere that is required to sustain a nuclear chain reaction. This provides a standardized basis for comparing the relative weaponization potential of different fissile materials.

The gold standard for determining critical masses come from Monte Carlo N-Particle (MCNP) transport simulations. Results are abundant in the literature. For pure 235U the bare mass is $46.4\pm1.7$ kg from MCNP calculations.\footnote{See \cite{chadwick2021nuclear} for a discussion of the history of this value.} 


\subsection{Bare Critical Masses}

The \textsc{FissionBomb} code from \cite{reed2009primer} and \cite{reed2010fission} is our starting point for approximating critical masses here and weapon yields in the next section. This code is simple and user friendly, eschewing all physics of energy-dependence in nuclear physics and assuming a pure 235U core. It is validated in \cite{reed2009primer} to obtain the yield of the Little Boy bomb from a numerical simulation of a roughly equivalent spherical core and tamper. Our 235U parameters in Tab. \ref{tab:1} predict a pure 235U bare mass of 48 kg, in fair agreement with Chadwick.

To adapt the \textsc{FissionBomb} code for lower enrichments, we follow some of the suggestions of \cite{lamoreaux2025elementary}. As \textsc{FissionBomb} was initially intended to model only the single nuclide 235U, the code takes minimal nuclear physics constants as input, including the fission cross section $\sigma_f$, the elastic scattering cross section $\sigma_{el}$, and $\nu$ the number of secondary neutrons per fission. To maintain this simplicity and avoid extensive code development, we build on top of its capabilities and obtain effective cross sections for the mixture of 235U and 238U. 

\begin{figure*}[t]
\includegraphics[width=0.99\textwidth]{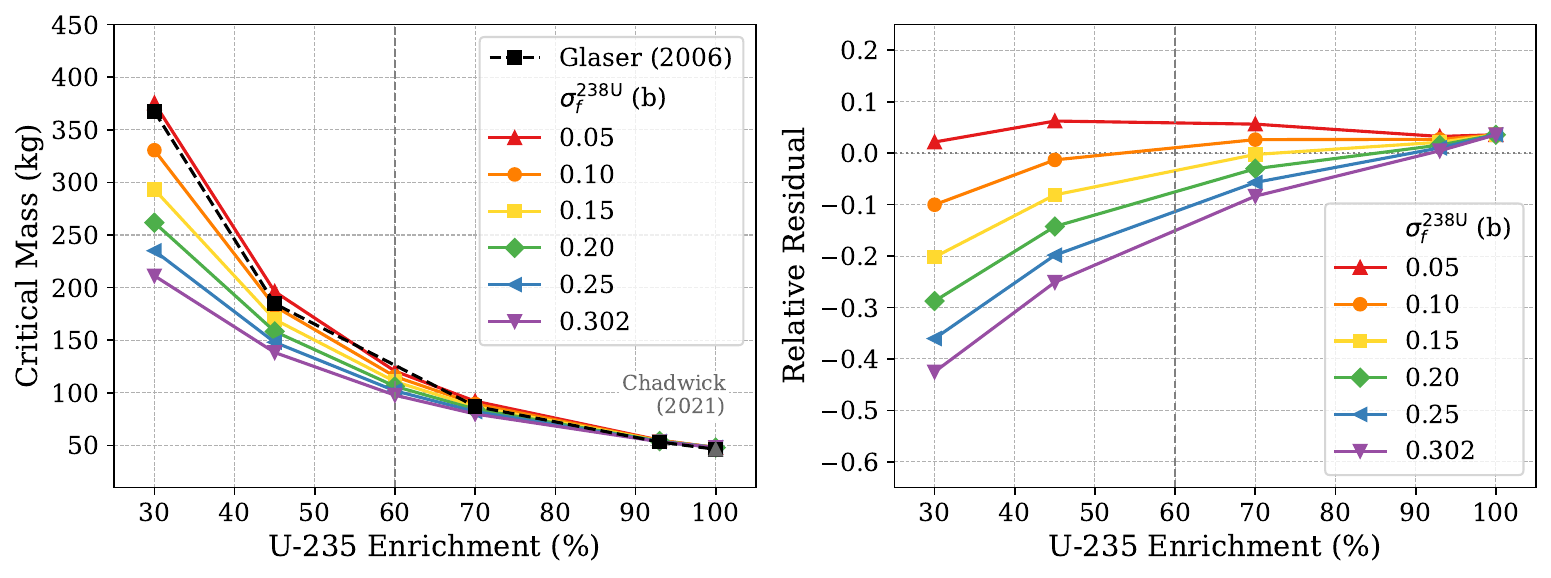}
\caption{\label{fig:barecrits_sigmaf} (Left) Bare critical mass dependence on $\sigma_f^{238\rm{U}}$
with MCNP values from \cite{glaser2006proliferation} and \cite{chadwick2021nuclear} and (right) residuals between our model calculations and the MCNP.}
\end{figure*}

First, \cite{lamoreaux2025elementary} suggests combining the elastic $\sigma_{el}$ and inelastic $\sigma_{in}$ scattering cross sections into a single transport cross section with the approximation ${\sigma_{tr} = \sigma_{el} + \frac{1}{2}\sigma_{in}}$. Fig. 1 of \cite{lamoreaux2025elementary} predicts a bare mass for pure 235U of 46 kg, in line with the modern value. 

For all other values of enrichment, we take a concentration weighted average of the 235U and 238U cross sections to obtain effective cross sections such that 

\begin{equation}
   \sigma_{el,\rm eff} = E_w \sigma_{tr}^{235\rm{U}} + (1-E_w) \sigma_{tr}^{238\rm{U}} 
\end{equation}

\noindent and

\begin{equation}
   \sigma_{f,\rm eff} = E_w \sigma_f^{235\rm{U}} + (1-E_w) \sigma_f^{238\rm{U}} 
\end{equation}

\noindent where $E_w$ is the enrichment percentage of 235U (as a decimal fraction). The atomic mass is treated similarly, and we use $\rho_{\rm U} = 19.05 \, \rm g \, cm^{-3}$ throughout. For simplicity, we take weight fractions and number fractions to be identical; this only introduces an error of order $10^{-2}$.

The choice of $\sigma_f^{238\rm{U}}$ is important, and we show this dependence in Fig. \ref{fig:barecrits_sigmaf}. Any constant value will notably neglect the energy dependence of the 238U fission cross section, which drops off sharply below about 1 MeV. As 1 MeV is a typical fission neutron energy, this means that after a few scatters neutrons become significantly less effective at causing 238U fission. In the absence of a true neutron spectrum, \cite{lamoreaux2025elementary} argues that lowering $\sigma_f^{238\rm{U}} = 0.302 \, \rm b$ to $\sigma_f^{238\rm{U}} = 0.10 \, \rm b$ captures the effect well. From Figure 1 in Lamoreaux, showing the critical mass as a function of enrichment for these two values, decreasing $\sigma_f^{238\rm{U}}$ this way increases the bare mass from 93 kg to 113 kg. A spline interpolation of the bare masses given in Table 1 of \cite{glaser2006proliferation} also gives a bare mass of 113 kg at 60\%-enrichment. 

The neutron multiplication is essential, and we make one final change to the \textsc{FissionBomb} code. For a composition average, unlike the collision and fission cross sections which are weighted by concentration, we also weight the average by the relative contribution of each isotope to the total fission cross section,
\begin{equation}
    \nu = E_w\frac{\sigma_f^{235\rm{U}}}{ \sigma_{f,\rm eff}} \nu^{235\rm{U}} + (1-E_w)\frac{\sigma_f^{238\rm{U}}}{ \sigma_{f,\rm eff}} \nu^{238\rm{U}}.
\end{equation}
\noindent The 235U always dominates the secondary neutrons. At the most extreme, $\nu = 2.46$ when using $\sigma_f^{238\rm U}=0.302$ at 30\%-enrichment. 

We show in Fig \ref{fig:barecrits_sigmaf} that ${\sigma_f^{238\rm{U}} = 0.10 \, \rm b}$ provides very good agreement, within 10\% of MCNP values at all enrichments. The best agreement at 60\%-enrichment is found at  ${0.10 \, \mathrm{b} \leq \sigma_f^{238\rm{U}} \leq 0.15 \, \rm b}$. 
The preferred $\sigma_f^{238\rm{U}}$ decreases with enrichment; this is unsurprising, as at high enrichment the 238U abundance is so small that it does not contribute to the fission yield, while at lower enrichments the 238U scattering cross section becomes the dominant cross section so the average energy of the neutrons will be lower. 

\textsc{FissionBomb} neglects the absorption cross section $\sigma_a$ and this disappearance reaction is a neutron sink. The absorption cross sections of 235U and 238U are similar and about $1\%$ of the total effective cross section, so we expect that \textsc{FissionBomb} is weakly overpredicting the number of neutrons. Though small, this can be important in determining yields, as the number of neutrons grows exponentially in a chain reaction and the final few generations of neutrons are responsible for the majority of energy output.  

To model a neutron sink, we experiment with an effective number of secondary fission neutrons $\nu_{\rm eff} = \eta \nu$ where $\eta$ is a dimensionless prefactor less than 1. This hack imagines that a neutron that is absorbed might as well have never been emitted for the purposes of the fission burn. However, as decreasing $\nu_{\rm eff}$ increases the critical mass, we commit to $\eta=1$ as our best parameter set with $\sigma_f^{\rm 238U} = 0.10 \, \rm b$ already weakly overpredicts the critical mass.

\subsection{Spontaneous Fission and Predetonation}

We remark here that a bare mass of 113 kg of 60\%-enriched uranium represents more than a quarter of the Iranian stockpile. Even if an aspiring nuclear terrorist could acquire that amount of fissile material, we argue that this is unsuitable for technical reasons. 

Predetonation, or a premature initiation of the chain reaction, is a critical failure mode in gun-type nuclear weapons. 
When detonated, the fissile material in a nuclear weapon is rapidly compressed or assembled to transition from a subcritical to a supercritical configuration, enabling the exponentially growing chain reaction \citep{reed2010predetonation}. However, assembly does not guarantee the availability of a first neutron to initiate the reaction; assembly simply reduces the mean free path of neutrons and allows the fission reaction to proceed efficiently. Predetonation occurs when a neutron, in our case originating from 238U spontaneous fission (SF), initiates the chain reaction prematurely while assembly is in progress. If the fissile material is still in a suboptimal state that is only weakly supercritical, the energy release causes the assembly to disassemble before reaching full yield, resulting in a significantly weaker ``fizzled'' detonation \citep{glaser2006proliferation,reed2011fizzle}. 

The 238U spontaneous fission rate is $10^3$ times greater than that of 235U, so the probability of predetonation scales strongly with enrichment for uranium weapons. Lower enrichment leaves more 238U in the core, increasing the spontaneous neutron background and, by extension, the risk of predetonation.  While an implosion type weapon is compressed over about a microsecond, in a gun-type weapon the insertion of the two subcritical pieces takes about a millisecond.

\begin{figure}[t]
\includegraphics[width=0.99\columnwidth]{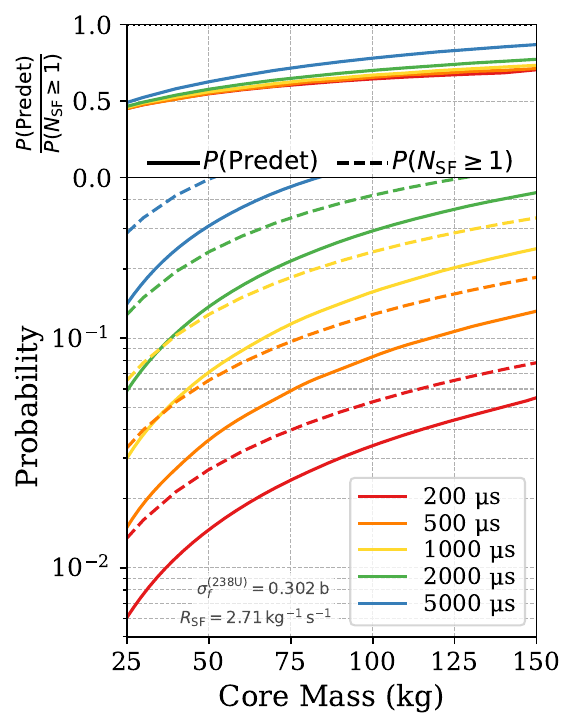}
\caption{\label{fig:predet} (Bottom) Probability of predetonation (solid) for different assembly times. These are lower than the probability of at least one spontaneous fission during an assembly time (dashed) because an emitted neutron must cause a fission. (Top) About half of spontaneous fissions can cause predetonation at masses below 50 kg.}
\end{figure}

Evaluating Poisson probabilities of decays allows weapon designers to assess the probability of a ``spontaneous-fission-free-millisecond'' and by extension whether or not the risk of predetonation is high. 
Using $R_{\rm SF}^{\rm 238U} = 6.78 \, {\rm  kg}^{-1} \, {\rm s}^{-1}$ we estimate a SF rate of $R_{\rm SF}^{\rm 60\%} = 2.71 \, {\rm  kg}^{-1} \, {\rm s}^{-1}$. Assuming a Poisson process, the probability of observing at least one decay in time interval $\Delta t$ is ${ P(N_{\rm SF} \geq 1) = 1 - \exp \left( {-R_{\rm SF} \Delta t} \right) }$.

For 113 kg at 60\%-enrichment, the SF rate is ${0.30 \, \rm ms^{-1}}$, so the risk of predetonation becomes high and thus a bare mass of 60\%-enriched uranium is unlikely to be suitable for an improvised bomb. Given the large investment of precious enriched uranium, it seems unlikely anyone would build a device so unlikely to function well. 

We present analytic results for the SF rates and detailed calculations of predetonation probabilities that incorporate fission and scattering cross sections in Fig. \ref{fig:predet} for assembly times spanning $200 \, \mu \rm s$ to $5000 \, \mu \rm s$. Our calculations for predetonation use the code from \cite{reed2010predetonation} which approximates the neutron transport in the core in addition to the Poissonian SF.
If our aspiring terrorists demand a successful explosion at full yield with a probability of at least 96\%, then their weapon must assemble no more than about 50 kg of 60\%-enriched uranium  in under $500 \, \mu \rm s$. Achieving an assembly time of $200 \, \mu \rm s$ is essential to reduce the predetonation probability to 1\% or lower. 
We also note that the ratio of predetonation probability to SF is mass dependent due to greater neutron escape from a smaller sphere. For masses below about 50 kg only about half of SFs during assembly can initiate predetonation. This analysis is largely insensitive to the exact value of $\sigma_f^{238\rm{U}}$, so we use the larger $\sigma_f^{238\rm{U}} = 0.302 \, \rm b$. This assumes that, following a SF, the 238U fission contribution to predetonation is dominated by a small integer numbers of unmoderated fission neutrons with MeV energies rather than a larger population that has scattered to lower energies like would be expected in a full chain reaction.

We emphasize that the methods of \cite{reed2010predetonation} assume a bare core for simplicity, so an oversized neutron reflecting tamper will increase the likelihood of predetonation slightly but certainly not to a value as great as the SF rate. A cautious reader could interpret the solid and dashed curves in Fig. \ref{fig:predet} as absolute upper limits and lower limits for predetonation. 

\begin{figure*}[t]
\includegraphics[width=0.99\textwidth]{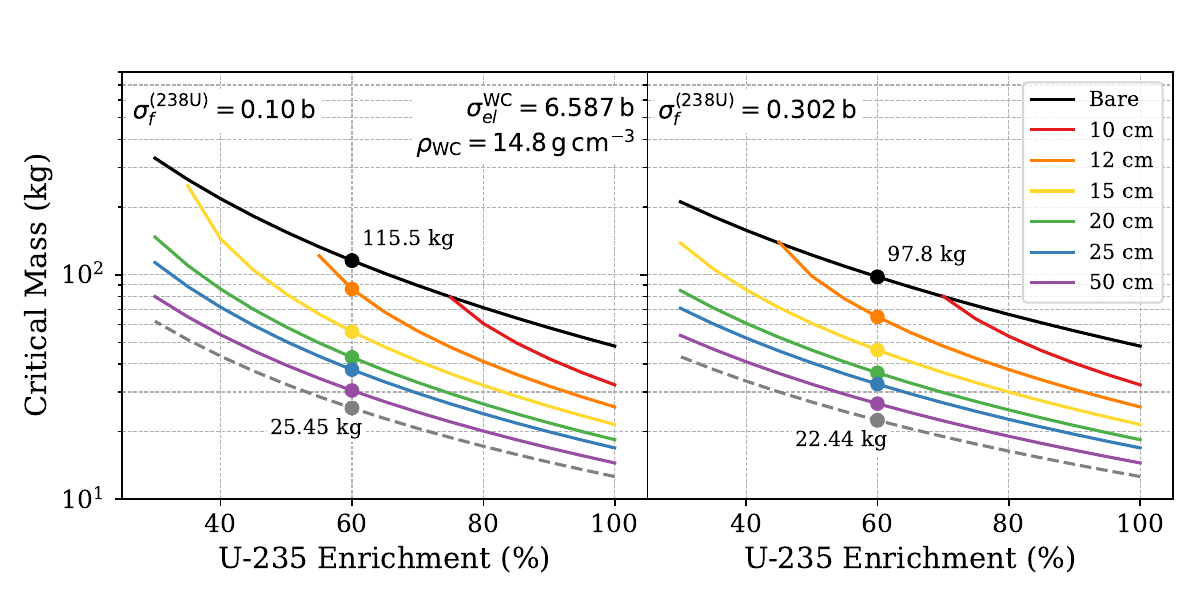}
\caption{\label{fig:tampedcrits} Critical mass dependence on WC tamper outer radius for $\sigma_f^{238\rm{U}} = 0.10 \, \rm b$  and $\sigma_f^{238\rm{U}} = 0.302 \, \rm b$. The dashed grey line at the bottom is the infinite tamper limit.}
\end{figure*}

Assembly times less than $500 \, \mu \rm s$ should not be especially challenging. Other works have shown that the gas pressures required to accelerate the core pieces together are comparable to those found in artillery, and do not exceed the yield strength of the kind of materials that would be used \citep[\textit{e.g.}][]{reed2014terrorist}. Those calculations are not repeated here, though it is worth noting that we consider much larger tamper radii and that may delay the assembly by a factor of a two relative to published estimates.

As a general observation, \citet{reed2010predetonation} estimates the predetonation likelihood for the Little Boy bomb with 64 kg of 80\%-enriched uranium and 200 $\mu \rm s$ assembly time to be between 1 and 2\%. A 32 kg core of 60\%-enriched uranium would contain an equal amount of 238U, and thus have similar failure rates, which is perhaps ideal for our aspiring terrorist. Though 32 kg is well below the bare mass, this could be made critical with a tamper.

\subsection{Tamped Critical Masses}

The aspiring nuclear terrorist circumvents the challenges posed by a bare critical mass and predetonation with a tamper, a layer of dense material surrounding the fissile core that reflects escaping neutrons back into the active volume and delays the disassembly of the core during detonation by inertial confinement, sustaining the chain reaction for a longer duration increasing the overall yield. Most importantly, a tamper reduces the critical mass of a fissile material by improving the neutron economy of the core. Neutron reflection effectively increases the neutron multiplication factor, allowing the weapon to achieve criticality with a smaller amount of fissile material. Together, these effects allow weapon designers to use less fissile material than would otherwise be required in a bare configuration.

While beryllium tampers are ideal, as discussed in other works, tungsten carbide (WC) is a more likely choice for an improvised device due to its availability and industrial applications. We use the default \textsc{FissionBomb} parameters for WC, $\rho = 14.8  \, \rm{g} \, \rm{cm}^{-3}$, $A=195.84$, and $\sigma_{el} = 6.587 \, \rm b$.\footnote{The density of pure WC is $15.6 \, \rm{g} \, \rm{cm}^{-3}$. The lower value used here is for cemented WC with a 6\% cobalt binder, a common industrially available form.}

In Fig. \ref{fig:tampedcrits} we show the tamped critical masses for a WC reflector of varying thicknesses for $\sigma_f^{238\rm{U}} = 0.10 \, \rm b$ and $\sigma_f^{238\rm{U}} = 0.302 \, \rm b$ for the same enrichments considered above. 
Because $\sigma_f^{238\rm{U}} = 0.10 \, \rm b$ weakly overpredicted the bare critical base and $\sigma_f^{238\rm{U}} = 0.302 \, \rm b$ underpredicted the bare critical mass at lower enrichments, the combination allows us to place an approximate bracket on the tamped critical mass at any given enrichment and tamper thickness. We expect that the true value is likely closer to what is predicted when using $\sigma_f^{238\rm{U}} = 0.10 \, \rm b$, and is perhaps 25 kg at 60\%-enrichment.

We remark on the missing data at low enrichment for small reflector thicknesses. The \textsc{FissionBomb} code solves for the tamped critical mass by solving the root finding problem presented in \cite{reed2009primer}. This required an approximate analytic treatment of a spherical core surrounded by a tamper shell of thickness $R_{\rm tamp} - R_{\rm core}$ with outer radius $R_{\rm tamp}$. This requires an appropriately chosen boundary condition for neutron leakage from the outer radius that can have numerical errors when the tamper thickness is similar to the neutron mean free path in the tamper, $\lambda_{\rm t} = 3.34 \, \rm cm$ for WC. Note that $R_{\rm core} = 8.6 \, \rm cm$ for a 50 kg core at 60\%-enrichment, close to the smallest $R_{\rm tamp}$ we consider. The numerical origin of this error is pronounced in some `near failure' cases where the tamped critical masses with 10 and 12 cm tampers numerically converged but exceed the bare critical masses by about a percent. 
This error ultimately does not affect our analysis much, as we can see numerical stability for tampers thicker than 15 cm at 60\%-enrichment.

When comparing the two cases in Fig. \ref{fig:tampedcrits}, a 10\% variation is observed at 60\% enrichment, which we take to be the error in our calculation of the tamped critical masses. 
Ultimately a 10\% uncertainty is not such a problem for this analysis. These plots show that a critical mass at 60\%-enrichment can be achieved with approximately 30 kg and a 50 cm tamper radius, all of which is achievable and unlikely to predetonate. If the aspiring terrorist is not able to steal many multiples of this mass to build two bombs, they will almost certainly use all of their uranium in the one weapon to maximize the yield, and we argued above that 50--60 kg core are also unlikely to predetonate. A 50--60 cm tamper is also efficient with uranium, having a tamped critical mass only 5 kg above the infinite tamper at 60\%-enrichment. For 30 to 70 kg cores, a 60 cm WC tamper is 13350 kg, well under shipping weight limits, and well under the 2.44 m width of a typical shipping container.

\section{Bomb Yields}\label{sec:bomb}

With our criticality analysis complete we now calculate the yields of weapons with various core masses and tamper thicknesses using the updated \textsc{FissionBomb} code. 

While a critical mass is achievable with about 25 kg, the yield will be unimpressive because the neutron multiplication rate $\alpha =  0$ at criticality because this condition is merely the threshold for a chain reaction to be self sustaining. To achieve the exponential growth of neutrons be to supercritical we require $\alpha>0$. 

The majority of the yield is released by the final few generations of neutrons due to the exponential growth of the chain reaction. The \textsc{FissionBomb} code handles this with a crude adaptive timestep, with one initial timestep to skip over the many neutron doubling times before the peak rate of fission $\phi_{\rm max}$. The initial timestep is chosen so that the fission rate is $\log[\phi / \rm (s^{-1}) ] = 20$  after the first step. Then, to resolve peak fission, we choose a timestep such that the fission rate rises to its peak and falls back to $10^{20} \, \rm s^{-1}$ over at least 1000 timesteps; $dt = 10^{-4} \, \mu \rm s$ is suitable for all core masses and tampers considered and yields robustly converge with this strategy.

We begin with a model of the Little Boy bomb to validate our effective nuclear physics. The Little Boy bomb model presented in \cite{reed2009primer} approximately reproduces the known yield of 12 to 15 kt by assuming a 64 kg spherical core of pure 235U with a 550 kg tamper of WC. The real Little Boy bomb however had a cylindrical 64 kg core of 80\%-enriched uranium and nested tamper with 300 kg of WC surrounded by 2000 kg of steel. Because \textsc{FissionBomb} requires a single composition for the tamper, we assume the steel is effectively worth only half its weight in WC and take a tamper mass of 1300 kg. This is not such a bad assumption; the typical density of steel is $\rm 7.75\textendash8.05 \, \rm g \, cm^{-3}$ and the angle integrated neutron elastic scattering cross section of iron is a few b near 1 MeV \citep{chen2020nuclear}. When using our effective cross sections and $\sigma_f^{238\rm{U}} = 0.10 \, \rm b$ ($\sigma_f^{238\rm{U}} = 0.15 \, \rm b$) we achieve a yield of 12.1 kt (12.8 kt), in line with the known value. It seems likely the over-estimation of the uranium purity and underestimation of the tamper mass produced canceling errors in \cite{reed2009primer}. Ultimately this approximation is most sensitive to the assumptions about the effective tamper mass; for every 100 kg added the yield rises by about 0.9 kt. Corrections for cylindrical compared to spherical geometry may also be of order 10\%.

\begin{figure}[t]
    \centering
    \includegraphics[width=0.48\textwidth]{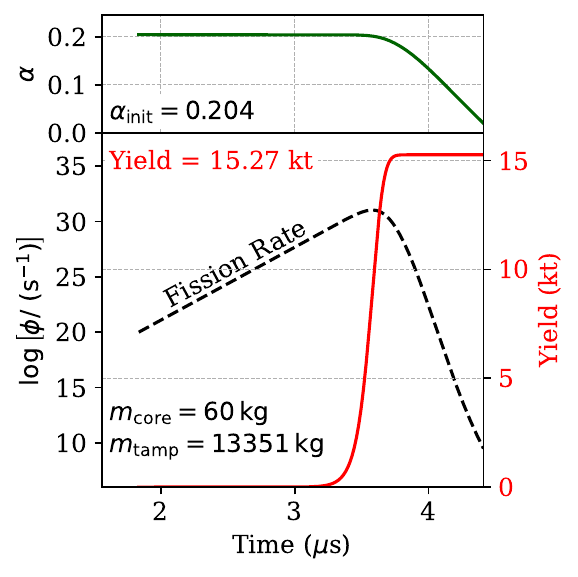}
    \caption{\label{fig:yieldtime} Fission burn simulation for a gun-type device with a 60 kg core surrounded by a 60 cm WC tamper (bottom). The majority of the yield is released within $0.15 \, \mu \rm s$, after which the core $\alpha$ (top) drops to zero due to the expansion of the core. Every explosion follows these qualitative curves.}
\end{figure}

We now consider a fiducial core of 60 kg of 60\%-enriched uranium with an approximate 60 cm tamper outer radius (13350 kg). In this case the neutron multiplication is more modest, $\alpha_{\rm init} = 0.204$ relative to $\alpha_{\rm LB} =0.388$ for our most conservative Little Boy models. 
In Fig. \ref{fig:yieldtime} we show a time dependent simulation of a fission chain reaction of a spherical mass and tamper that reaches a yield of 15.27 kilotons. Due to the lower neutron multiplication, this device takes much longer to release its yield, over $4 \, \mu \rm s$ compared to $1.5 \, \mu \rm s$ for the simulated Little Boy above. 
Our yield uncertainty may also be 10\%, and with our conservative assumptions for nuclear parameters we expect that we systematically underpredict yields. This device is remarkably successful, and the core contains 36 kg of 235U compared to Little Boy's 51.2 kg but achieves a comparable yield due to the oversized tamper.

The fission burn in Fig. \ref{fig:yieldtime} reaches a peak fission rate of $\log[\phi_{\rm max} / \rm (s^{-1}) ] = 31.038$ at $3.58 \, \mu \rm s$, with 59\% of the total yield released.  When at 99\% of the full yield, the core density has only decreased from $19.05 \, \rm{g \, cm}^{-3}$ to $18.426 \, \rm{g \, cm}^{-3}$, but the fission rate is still $\log[\phi / \rm (s^{-1}) ] = 29.967$. Half of the yield is released over about a thousand timesteps ${(\sim 0.15 \, \mu \rm s)}$ around peak fission. 

Changes in anything that affects neutron multiplication, such as fission cross sections and neutrons per fission, should have exponential effects on the energy release by way of their affect on the initial $\alpha$. For comparison, using $\sigma_f^{238\rm{U}} = 0.15 \, \rm b$ we find an initial $\alpha = 0.216$, so one expects the time to peak fission to occur after $(0.204/0.216) \times 3.57 \, \mu \rm{s} = 3.38 \, \mu \rm{s}$. A simulation finds nearly that, but the yield rises to 18.36 kt with a greater peak fission rate of $\log[\phi_{\rm max} / \rm (s^{-1}) ] = 31.154$. Longer fission burns therefore tend to lower the yield and $\phi_{\rm max}$, as the core has had more time to expand. 
Taken together, these two simulations might imply an expected 17 kt yield with 2 kt uncertainty from nuclear physics parameters. Note that yields are insensitive to the assumed energy per fission or energy per fission neutron. The core expansion rate is responsible for quenching the fission rate and is largely a function of yield and pressure, so increasing the energy released per fission simply reduces the peak fission rate with the yield remaining relatively constant.

\begin{figure}[t]
    \centering
    \includegraphics[width=0.48\textwidth]{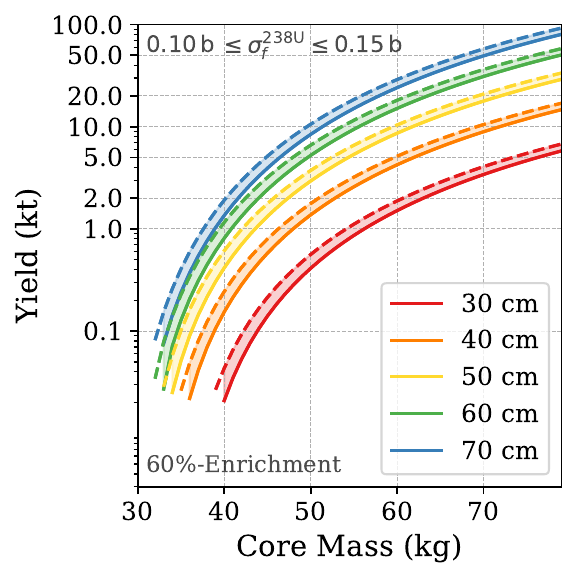}
    \caption{\label{fig:yieldvstamp} Yields for a range of core masses and tamper outer radii. Solid lower bounds (dashed upper bounds) use $\sigma_f^{238\rm{U}} = 0.10 \, \rm b$ ($\sigma_f^{238\rm{U}} = 0.15 \, \rm b$). Legend corresponds to outer tamper radii. Each radius actually corresponds to a fixed tamper mass assuming a 50 kg core ($R_{\rm core} = 8.557 \, \rm cm$), so outer radii are not exact. See Tab.~\ref{tab:yieldsum} for data.}
\end{figure}

In Fig. \ref{fig:yieldvstamp} we consider how limitations of available uranium and tamper mass translate to limits on the yield and estimate the minimum quantity of each to achieve a yield worth the investment, perhaps 1 kt. The colored bands are bounded by assuming $\sigma_f^{\rm 238U} = 0.10 \, \rm b$ (lower limit, solid) and $\sigma_f^{\rm 238U} = 0.15 \, \rm b$ (upper limit, dashed). We use fixed tamper masses of 1634.3 kg, 3928.1 kg, 7709.7 kg, 13351.2 kg, and 21224.4 kg, so the outer tamper radii reported in the legend are only exact for a 50 kg core but are otherwise still good to better than 1\%.
The typical cargo limit for a standard twenty foot container is 21600 kg, so a 70 cm tamper strains standard delivery limits and is taken as an absolute upper limit on the tamper size.\footnote{Heavy tested containers have cargo limits of 28080 kg.}

\begin{table*}[t]
\centering
\renewcommand{\arraystretch}{1.1}
\resizebox{\textwidth}{!}{%
\begin{tabular}{c|c}
\begin{minipage}{0.48\textwidth}
\centering
\textbf{Yield (kiloton TNT equivalent)}\\[3pt]
\begin{tabular}{c|cccccc}
\hline
$R_{\rm tamp}$ / $M_{\rm core}$  & 35 kg & 40 kg & 50 kg & 60 kg & 70 kg & 80 kg \\
\hline
30 cm &       &     & 0.41 & 1.53 & 3.46 & 6.20 \\
40 cm &      & 0.15 & 1.38 & 4.29 & 9.05 & 15.67 \\
50 cm & 0.05  & 0.42 & 2.96 & 8.73 & 18.07 & 30.98 \\
60 cm & 0.13  & 0.81 & 5.28 & 15.27 & 31.43 & 53.70 \\
70 cm & 0.24  & 1.36 & 8.47 & 24.36 & 50.02 & 85.35 \\
\hline
\end{tabular}
\end{minipage}
&
\begin{minipage}{0.48\textwidth}
\centering
\textbf{Peak Fission Rate $\log[\phi / \rm (s^{-1}) ] $}\\[3pt]
\begin{tabular}{c|cccccc}
\hline
$R_{\rm tamp}$ / $M_{\rm core}$  & 35 kg & 40 kg & 50 kg & 60 kg & 70 kg & 80 kg \\
\hline
30 cm &       &       & 29.26 & 30.02 & 30.50 & 30.84 \\
40 cm &       & 28.52 & 29.82 & 30.48 & 30.92 & 31.24 \\
50 cm & 27.74 & 28.98 & 30.15 & 30.80 & 31.22 & 31.54 \\
60 cm & 28.18 & 29.29 & 30.41 & 31.04 & 31.46 & 31.78 \\
70 cm & 28.47 & 29.51 & 30.61 & 31.24 & 31.67 & 31.98 \\
\hline
\end{tabular}
\end{minipage}
\\[12ex]
\begin{minipage}{0.48\textwidth}
\centering
\textbf{$\alpha_{\rm init}$}\\[3pt]
\begin{tabular}{c|cccccc}
\hline
$R_{\rm tamp}$ / $M_{\rm core}$  & 35 kg & 40 kg & 50 kg & 60 kg & 70 kg & 80 kg \\
\hline
30 cm &       &       & 0.124 & 0.196 & 0.259 & 0.315 \\
40 cm &       & 0.061 & 0.135 & 0.203 & 0.264 & 0.319 \\
50 cm & 0.030 & 0.066 & 0.137 & 0.204 & 0.265 & 0.320 \\
60 cm & 0.034 & 0.068 & 0.138 & 0.204 & 0.265 & 0.320 \\
70 cm & 0.035 & 0.068 & 0.138 & 0.204 & 0.265 & 0.320 \\
\hline
\end{tabular}
\end{minipage}
&
\begin{minipage}{0.48\textwidth}
\centering
\textbf{Peak Fission Time ($\mu \rm s$)}\\[3pt]
\begin{tabular}{c|cccccc}
\hline
$R_{\rm tamp}$ / $M_{\rm core}$  & 35 kg & 40 kg & 50 kg & 60 kg & 70 kg & 80 kg \\
\hline
30 cm &       &       & 5.46 & 3.58 & 2.76 & 2.30 \\
40 cm &       & 10.78 & 5.15 & 3.52 & 2.75 & 2.31 \\
50 cm & 20.70 & 10.15 & 5.14 & 3.55 & 2.78 & 2.34 \\
60 cm & 19.13 & 10.02 & 5.17 & 3.58 & 2.81 & 2.36 \\
70 cm & 18.62 & 10.03 & 5.21 & 3.61 & 2.83 & 2.38 \\
\hline
\end{tabular}
\end{minipage}
\\[12ex]
\begin{minipage}{0.48\textwidth}
\centering
\textbf{Fission Efficiency (\%)}\\[3pt]
\begin{tabular}{c|cccccc}
\hline
$R_{\rm tamp}$ / $M_{\rm core}$  & 35 kg & 40 kg & 50 kg & 60 kg & 70 kg & 80 kg \\
\hline
30 cm &       &       & 0.04  & 0.13  & 0.25  & 0.40  \\
40 cm &       & 0.02  & 0.14  & 0.37  & 0.66  & 1.00  \\
50 cm & 0.01  & 0.05  & 0.30  & 0.75  & 1.32  & 1.98  \\
60 cm & 0.02  & 0.10  & 0.54  & 1.30  & 2.30  & 3.44  \\
70 cm & 0.04  & 0.17  & 0.87  & 2.08  & 3.66  & 5.47  \\

\hline
\end{tabular}
\end{minipage}
&
\begin{minipage}{0.48\textwidth}
\centering
\textbf{Time at 99\% of Yield ($\mu \rm s$)}\\[3pt]
\begin{tabular}{c|cccccc}
\hline
$R_{\rm tamp}$ / $M_{\rm core}$  & 35 kg & 40 kg & 50 kg & 60 kg & 70 kg & 80 kg \\
\hline
30 cm &       &       & 5.69  & 3.73  & 2.87  & 2.39  \\
40 cm &       & 11.26 & 5.37  & 3.67  & 2.87  & 2.40  \\
50 cm & 21.65 & 10.59 & 5.35  & 3.69  & 2.89  & 2.43  \\
60 cm & 19.99 & 10.46 & 5.39  & 3.73  & 2.92  & 2.45  \\
70 cm & 19.45 & 10.46 & 5.43  & 3.76  & 2.94  & 2.47  \\
\hline
\end{tabular}
\end{minipage}
\end{tabular}
}
\caption{\label{tab:yieldsum} Yields, $\alpha_{\rm init}$, fission efficiency, peak fission rate, and time of peak fission, and time for the fission burn to reach 99\% of full yield for select simulations in Fig.~\ref{fig:yieldvstamp} with $\sigma_f^{238\rm{U}} = 0.10 \, \rm b$. At peak fission the yield is 58--59\% of full yield.}
\end{table*}

If an aspiring terrorist desires a yield of at least 1 kt, only about 40 kg of 60\%-enriched uranium is required. Yields also grow rapidly with core mass. The benefits of the oversized tamper are apparent, with an order of magnitude growth in yields going from ${R_{\rm tamp} = 30 \, \rm cm}$ to ${R_{\rm tamp} = 60 \, \rm cm}$ for fixed core mass. With only 38 kg of uranium a 70 cm tamper may suffice. 
Note that yield predictions below 0.1 kt may be suspect in this model. The tamper mass is approximately 10--20 tonnes, so yields of this order may be contained by the tamper, with the explosion only shattering and dispersing it. This is not a problem for this analysis, as we consider it unlikely that an aspiring nuclear terrorist would be content with yields lower than 1 kt.
Such a yield would be comparable to the Beirut port explosion of 2020, and while still devastating, it could be achieved much more easily by loading a ship with a few thousand tonnes of ammonium nitrate. Such a conventional attack would also be much easier to disguise as regular commerce, and less likely to attract attention.

The maximum yield could approach 100 kt, but this strains the limits of delivery mass and has higher risks of predetonation, especially as each of the two core pieces become comparable to a tamped critical mass. Any larger than this and a device may require both core halves to be launched into the tamper, complicating the device and creating more opportunities for the weapon to fail. 

We also note that a shortage of tungsten may not limit the yields as much as Fig. \ref{fig:yieldvstamp} predicts if an outer tamper made of steel is added. The neutron economy benefits most strongly from the innermost tamper being tungsten, as in the Little Boy bomb simulated earlier. Future work developing \textsc{FissionBomb} to model nested tampers is needed to explore this in detail.

Lastly, we note that gun-type weapons use uranium inefficiently. Famously, only about 1\% of the uranium in the Little Boy bomb underwent fission, with the bomb disassembling and dispersing the rest. The most massive cores simulated here can reach yields of approximately 1 kt per kg of core mass (see Tab. \ref{tab:yieldsum}) implying efficiencies of about 5\%. Such low fission rates justify many of the assumptions in the \textsc{FissionBomb} code, such as a neglecting the declining abundance of uranium and the rising presence of fission fragments that could act as neutron scatterers and absorbers. Future iterations of the \textsc{FissionBomb} code could self consistently track the nuclear abundances, but we expect the impact on yields to be small, lowering them by a factor no larger than the efficiency found here. Neutron absorption in the tamper is worth investigating given the large size.


\section{Summary\label{sec:sum}}

For the first time in decades, the prospect of a non-state or sub-state actor acquiring fissile material is no longer a distant hypothetical but a pressing and immediate threat. Historically, enrichment and the challenge of obtaining fissile material has been the greatest barrier to a weapon. While we have not considered other technical challenges an aspiring nuclear terrorist might face when building a nuclear weapon, such as uranium conversion, material acquisition, or delivery, these issues are discussed extensively elsewhere and do not pose as significant of a barrier as the acquisition of enriched uranium.

We have demonstrated that with a sufficiently large neutron reflecting tamper 60\%-enriched uranium can be used to construct a kiloton-yield nuclear weapon with as little as 40 kg of fissile material. Such a weapon would be relatively simple, and following the design of the Little Boy bomb it could be expected to successfully detonate without testing. Though too large and massive to be delivered by a ballistic missile, such a weapon could be smuggled in a standard shipping container on a truck or by cargo ship into a port.

\acknowledgements

The authors thank Cameron Reed for his codes and discussion. We also thank Vesal Razavimaleki and Matthias Perdekamp for conversation. Financial support for this publication comes from Cottrell Scholar Award \#CS-CSA-2023-139 sponsored by Research Corporation for Science Advancement.


\bibliographystyle{apsrev4-1}
\bibliography{references}

\end{document}